\documentclass[%
 aip,
 8pt,
 amsmath,amssymb,
reprint,%
]{revtex4-1}

\usepackage{amsmath}
\usepackage{multirow}
\usepackage[utf8]{inputenc}                         
\usepackage{amsthm}                                  
\usepackage{amscd}                                    
\usepackage{amssymb}                                  
\usepackage{verbatim}                                 
\usepackage{breqn}
\usepackage{graphics}                            
\usepackage{subfigure}   
\usepackage{stackengine}           
\usepackage{bm}                                        
\usepackage{float} 
\usepackage{color} 
\usepackage{textcomp}
\usepackage{stmaryrd}
\usepackage{epstopdf}
\usepackage{mathrsfs}
\usepackage{calligra}
\usepackage{url}
\usepackage{breqn}
\usepackage{enumitem} 
\usepackage{booktabs}
\usepackage[colorlinks,linkcolor=red,citecolor=red,urlcolor=blue]{hyperref}% add hypertext 

\usepackage{array}
\newcolumntype{L}[1]{>{\raggedright\let\newline\\\arraybackslash\hspace{0pt}}m{#1}}
\newcolumntype{C}[1]{>{\centering\let\newline\\\arraybackslash\hspace{0pt}}m{#1}}
\newcolumntype{R}[1]{>{\raggedleft\let\newline\\\arraybackslash\hspace{0pt}}m{#1}}

\usepackage{graphicx}% Include figure files
\usepackage{dcolumn}% Align table columns on decimal point
\usepackage{bm}% bold math
\usepackage{hyperref}% add hypertext capabilities
\usepackage[mathlines]{lineno}% Enable numbering of text and display math
\begin{document}

\title{A distribution-free lattice Boltzmann method for compartmental reaction–diffusion systems with application to epidemic modelling}

\author{Alessandro De Rosis}
 \email{alessandro.derosis@manchester.ac.uk}
\affiliation{Department of Mechanical and Aerospace Engineering\\
The University of Manchester$,$ Manchester$,$  M13 9PL United Kingdom}

\date{\today}

\begin{abstract}
We introduce a distribution-free lattice Boltzmann formulation for general compartmental reaction--diffusion systems arising in mathematical epidemiology. The proposed scheme, termed a single-step simplified lattice Boltzmann method (SSLBM), evolves directly macroscopic compartment densities, eliminating the need for particle distribution functions and explicit streaming operations. This yields a compact and computationally efficient framework while retaining the kinetic consistency of lattice Boltzmann methodologies.

\noindent The approach is applied to a SEIRD (Susceptible--Exposed--Infected--Recovered--Deceased) reaction--diffusion model as a representative case. The resulting discrete evolution equations are derived and shown to recover the target macroscopic dynamics. The method is systematically validated against a fourth-order finite difference reference solution and compared with a standard BGK lattice Boltzmann formulation.

\noindent Numerical results demonstrate that the SSLBM consistently improves accuracy across all compartments and norms. The error reduction is robust with respect to both the basic reproduction number and diffusion strength, typically ranging between factors of approximately two and five depending on the regime. In particular, the method shows enhanced control of localised errors in regimes characterised by strong nonlinear coupling and steep spatial gradients. Our findings indicate that the proposed formulation provides an accurate and efficient alternative to classical lattice Boltzmann approaches for reaction--diffusion systems, with particular advantages in stiff and nonlinear epidemic dynamics.
\end{abstract}

\maketitle

% -----------------------------------------------------------------------
% INTRODUCTION
% -----------------------------------------------------------------------
\section{Introduction}
The mathematical modelling of epidemic dynamics has a long and productive history and has
been successfully applied to diverse outbreak scenarios, including
SARS~\cite{dye2003modeling} and
Ebola~\cite{https://doi.org/10.1111/j.1541-0420.2006.00609.x}. The most widely adopted
framework is the compartmental approach, in which the total population is partitioned into
distinct groups according to disease status, and the flow of individuals between groups is
governed by biologically motivated transition rules~\cite{brauer2019mathematical}. The
foundational work of Kermack and McKendrick~\cite{doi:10.1098/rspa.1927.0118} established
a remarkably general integro-differential framework for epidemic modelling; the
susceptible--infected--recovered (SIR) ordinary differential equation system, while often
attributed to that paper, is in fact a special limiting case of their broader
formulation. A modern treatment of its derivation and
implications can be found in the work by Breda \textit{et al.}~\cite{breda2012formulation}.\\
\indent Building on the SIR framework, richer compartmental structures have been developed to
account for additional stages of disease progression. We consider here one specific
instance, the susceptible--exposed--infected--recovered--deceased (SEIRD) structure, which
extends SIR by introducing an exposed ($E$) compartment for individuals who have been
infected but are not yet infectious, and a deceased ($D$) compartment to track
fatalities~\cite{korolev2021identification}. Further refinements of compartmental
structures, including vaccination coverage, maternal immunity, and age-stratified
structure~\cite{doi:10.1080/17477778.2021.1977731}, can enable increasingly realistic
descriptions of epidemic dynamics.\\
\indent Despite their widespread use, ordinary differential equation (ODE)-based
modelling is inherently non-spatial: it treats the entire population as
perfectly mixed and therefore cannot capture the heterogeneous spatial transmission
patterns that often govern real epidemic
outbreaks~\cite{gatto2020spread,liu2007formation}. Partial differential equation
(PDE)-based formulations overcome this limitation by augmenting the compartmental
equations with spatial operators, enabling explicit representation of disease spreading.
The simplest such operator is Fickian diffusion, which models the random dispersal of
individuals across space. More complex spatial dynamics, including nonlinear or
spatially varying diffusion, advective transport, and chemotactic-like
terms~\cite{ramaswamy2021comprehensive},are also possible and have been studied in
recent years~\cite{bertaglia2021hyperbolic, colli2021seird}. Although PDE models impose a
greater computational burden, they naturally accommodate geographical heterogeneity,
population density variations, and multiscale
interactions~\cite{doi:10.1142/S0218202520500323}, making them valuable for spatially
resolved epidemic forecasting.\\
\indent The lattice Boltzmann method (LBM) offers an alternative discretisation strategy
well suited to reaction-diffusion systems. Rather than explicitly computing the Laplacian
operators that appear in PDE-based epidemic models, the LBM evaluates diffusion implicitly
through a streaming-and-collision cycle, with the diffusivity encoded through a single
relaxation parameter. In a previous study~\cite{PhysRevE.102.023301}, we demonstrated
that a Bhatnagar--Gross--Krook (BGK) lattice Boltzmann scheme provides accurate and efficient solutions of the spatial SIR structure. The present work extends that approach to the more complex SEIRD system and, critically, introduces a single-step simplified lattice Boltzmann method (SSLBM) that eliminates the storage of particle distribution functions
(PDFs) altogether. The SSLBM derivation is cast in terms of a generic compartmental
reaction-diffusion PDE, so that the resulting algorithm applies immediately to any
compartmental epidemic model of this class. The SSLBM builds on the simplified LBM (SLBM)
framework of Refs.~\cite{chen2017simplified,chen2018high,chen2018improvements} and the
single-step strategy of Delgado-Guti\'{e}rrez \textit{et
al.}~\cite{delgado2021single}, which has previously been validated for
magnetohydrodynamics~\cite{10.1063/5.0058884} and shallow water
flows~\cite{10.1063/5.0147175} but not yet for reaction-diffusion systems. The goals of
this study are: (i) to derive the SSLBM for a generic compartmental PDE and apply it to
the SEIRD system; (ii) to formally analyse its consistency, stability, and mass
conservation; and (iii) to validate it against a fourth-order finite differences discretisation and benchmark it against the BGK LBM.

% -----------------------------------------------------------------------
% THE SEIRD MODEL
% -----------------------------------------------------------------------
\section{Methodology}
In this section, we first introduce the SEIRD reaction–diffusion system and its formulation in a generic compartmental framework. We then outline its kinetic discretisation through the BGK lattice Boltzmann method and present its simplified and single-pass reformulations leading to the SSLBM. Finally, we show how the resulting scheme replaces explicit spatial derivatives with local lattice operations while preserving consistency with the underlying kinetic description and conservation properties.
\subsection{Macroscopic governing equations}
By assuming a two-dimensional Cartesian reference system of axes $x-y$, we consider the following specific SEIRD reaction-diffusion
system~\cite{VIGUERIE2021106617}:
\begin{eqnarray}
\partial_t S &=& -\frac{\beta S I}{N_c} + d^{S} \nabla^2 S, \label{EqSEIRD_S} \\
\partial_t E &=& \frac{\beta S I}{N_c} - \alpha E + d^{E} \nabla^2 E, \label{EqSEIRD_E}\\
\partial_t I &=& \alpha E - \gamma I + d^{I} \nabla^2 I, \label{EqSEIRD_I}\\
\partial_t R &=& \left(1-\phi \right) \gamma I + d^{R} \nabla^2 R, \label{EqSEIRD_R}\\
\partial_t D &=& \phi \gamma I, \label{EqSEIRD_D}
\end{eqnarray}
where $\partial_t$ denotes the partial time derivative and $\nabla^2 =
\partial_{xx}+\partial_{yy}$ is the two-dimensional Laplacian. The dependence of the compartments on the spatial coordinate $\mathbf{x}$ and time $t$ is implicitly assumed. The total population is $N = S+E+I+R+D$, but the force-of-infection term uses only the \emph{contact-eligible}
population $N_c = S+E+I+R$, which excludes the deceased since dead individuals do not
participate in disease transmission. The epidemiological parameters and their units are
summarised in Table~\ref{Table_params}. The diffusion coefficients $d^{S}$, $d^{E}$,
$d^{I}$, $d^{R}$ (units: $\mathrm{km}^2\,\mathrm{day}^{-1}$) quantify the spatial
mobility of each compartment; the deceased do not diffuse.
\begin{table*}[!htbp]
\centering
\begin{tabular}{c|l|c}
\hline\hline
Parameter & Description & Units \\
\hline
$\beta$   & Contact rate (times probability of infection per contact) & $\mathrm{day}^{-1}$ \\
$\alpha$  & Inverse of the mean incubation period & $\mathrm{day}^{-1}$ \\
$\gamma$  & Removal rate (transition rate from $I$ to $R$ or $D$) & $\mathrm{day}^{-1}$ \\
$\phi$    & Infection fatality ratio & --- \\
$d^C$     & Diffusion coefficient for compartment $C\in\{S,E,I,R\}$ & $\mathrm{km}^2\,\mathrm{day}^{-1}$ \\
\hline\hline
\end{tabular}
\caption{Epidemiological parameters appearing in the SEIRD compartmental structure
Eqs.~(\ref{EqSEIRD_S}--\ref{EqSEIRD_D}), with descriptions and units. Note that $\beta$
is often called the contact rate and combines the rate of contacts with the per-contact
transmission probability; $\gamma$ is the removal rate governing transition from the
infectious compartment to either recovery or death.}
\label{Table_params}
\end{table*}

\indent Equations~(\ref{EqSEIRD_S}--\ref{EqSEIRD_R}) form a coupled system of
reaction-diffusion equations, with the deceased compartment evolving purely in time
according to the ODE~(\ref{EqSEIRD_D}). Direct numerical solution via finite differences
requires the evaluation of four Laplacian operators per grid point and time step, which can represent a significant fraction of the total computational cost. The goal of this work is to develop a more efficient numerical approach that avoids explicit computation of these operators.

% -----------------------------------------------------------------------
% GENERIC COMPARTMENTAL PDE AND THE SSLBM
% -----------------------------------------------------------------------
\indent Before introducing the specific SEIRD discretisation, we derive the SSLBM at the
level of a \emph{generic} compartmental PDE. This generality is intentional: the
derivation applies to any compartmental epidemic model in which each compartment $C$
evolves according to a reaction-diffusion equation of the form
\begin{equation}
\partial_t C = \Psi^C + d^C \nabla^2 C,
\label{generic_pde}
\end{equation}
where $\Psi^C$ is the net reaction source term (which may couple $C$ to other
compartments) and $d^C \ge 0$ is the diffusion coefficient. The SEIRD
system~(\ref{EqSEIRD_S}--\ref{EqSEIRD_R}) is a specific instance of
Eq.~(\ref{generic_pde}) with four coupled compartments. 

\subsection{BGK lattice Boltzmann method}
The BGK lattice Boltzmann method solves Eq.~(\ref{generic_pde}) by associating each
compartment $C$ with a set of particle distribution functions $f_i^C$, where $i =
0,\ldots,4$ indexes the discrete velocity directions of the D2Q5 lattice. The link
vectors of this lattice are $\mathbf{c}_i = [c_{ix}, c_{iy}]$ with
\begin{equation}
c_{ix} = [0,\,1,\,0,\,-1,\,0], \qquad
c_{iy} = [0,\,0,\,1,\,0,\,-1],
\end{equation}
and the corresponding weights are $w_0 = 1/3$, $w_{1,2,3,4} =
1/6$~\cite{KRUGER_Book_2017}. The PDFs obey the BGK lattice Boltzmann equation (LBE)
\begin{widetext}
\begin{equation}\label{pdf}
f_i^C(\mathbf{x}+\mathbf{c}_i \Delta t,\,t+\Delta t)
= f_i^C(\mathbf{x},t)
+ \frac{1}{\tau^C}\!\left[ f_i^{C,\mathrm{eq}}(\mathbf{x},t) - f_i^C(\mathbf{x},t) \right]
+ w_i \Psi^C(\mathbf{x},t) \Delta t,
\end{equation}
\end{widetext}
where time and space are expressed in lattice Boltzmann units with $\Delta t = 1$ (in lattice units). The equilibrium distribution is $f_i^{C,\mathrm{eq}} = w_i \rho^C$, where the
macroscopic compartment density $\rho^C = \sum_i f_i^C$ plays the role of the population
density of compartment $C$. The relaxation time $\tau^C$ is linked to the diffusivity
by~\cite{zhang2012lattice}
\begin{equation}
d^C = c_s^2\!\left(\tau^C - \tfrac{1}{2}\right),
\label{diffusivity}
\end{equation}
where $c_s = 1/\sqrt{3}$ is the lattice speed of sound~\cite{doi:10.1063/1.464316}. The BGK LBM
inherently accounts for spatial diffusion, without requiring any explicit Laplacian evaluation. For the SEIRD system, the four reactive source terms are
\begin{eqnarray}
\Psi^{S} &=& -\frac{\beta \rho^S \rho^I}{\rho^{N_c}}, \nonumber \\
\Psi^{E} &=& \frac{\beta \rho^S \rho^I}{\rho^{N_c}} - \alpha \rho^E, \nonumber \\
\Psi^{I} &=& \alpha \rho^E - \gamma \rho^I, \nonumber \\
\Psi^{R} &=& \left(1-\phi \right) \gamma \rho^I,
\label{sources}
\end{eqnarray}
where $\rho^{N_c} = \rho^S + \rho^E + \rho^I + \rho^R$ is the density of the
contact-eligible population.

\indent For $M$ total grid points, the BGK LBM requires storing five PDFs per compartment
for four mobile groups, together with $\rho^D$ at two consecutive time levels, giving a
minimum storage of $22M$ floating-point values. By contrast, a FDM implementation of the
same equations needs only $10M$ values, retaining only the macroscopic densities. This
disparity motivates the development of PDF-free methods that preserve LBM accuracy at FDM
memory cost.

\subsection{Simplified lattice Boltzmann method}
The simplified LBM (SLBM) of Chen \textit{et al.}~\cite{chen2017simplified,
chen2018high, chen2018improvements} achieves this by evolving only macroscopic densities
via a predictor--corrector sequence:

\begin{align}
\rho^{C,\star}\!\left(\mathbf{x}, t+\Delta t \right)
&= \sum_i w_i\, \rho^C\!\left(\mathbf{x}-\mathbf{c}_i \Delta t,\,t \right),
\label{SLBM1}\\
\rho^C\!\left(\mathbf{x}, t+\Delta t \right)
&= \rho^{C,\star}\!\left(\mathbf{x}, t+\Delta t \right) + \nonumber \\
& \left(\tau^C-1\right) \!\Bigg[
\sum_i w_i\, \rho^{C,\star}\!\left(\mathbf{x}+\mathbf{c}_i \Delta t,\,t+\Delta t \right)-  \nonumber \\
&\qquad \qquad \rho^C\!\left(\mathbf{x}, t\right) \Bigg]  + \Psi^C\!\left(\mathbf{x}, t\right)\Delta t.
\label{SLBM2}
\end{align}
The SLBM requires $10M$ values (two densities per compartment per grid point, plus $\rho^D$
at two levels), matching the FDM storage. However, its predictor--corrector structure
visits each grid point twice per time step, partially offsetting the memory advantage.\\
\indent To eliminate the double-pass overhead, Delgado-Guti\'{e}rrez \textit{et
al.}~\cite{delgado2021single} proposed the SSLBM, in which each node is processed exactly
once. We derive it here for the generic compartment $C$, working from the BGK LBE without
source term:
\begin{equation}
f_i(\mathbf{x}+\mathbf{c}_i \Delta t,\, t+\Delta t)
= f_i(\mathbf{x}, t)
+ \frac{1}{\tau}
\left[ f_i^{\mathrm{eq}}(\mathbf{x}, t) - f_i(\mathbf{x}, t) \right].
\label{lbe_clean}
\end{equation}
We decompose $f_i = f_i^{\mathrm{eq}} + f_i^{\mathrm{neq}}$ with $f_i^{\mathrm{eq}} = w_i
\rho$. A standard Chapman--Enskog multiscale
expansion~\cite{KRUGER_Book_2017} of Eq.~(\ref{lbe_clean}),in which the
distribution function is expanded in powers of a small Knudsen number and time/space
derivatives are ordered by scale, yields, at leading order, the non-equilibrium
contribution
\begin{equation}
f_i^{\mathrm{neq}}
= -\tau \left( \partial_t + \mathbf{c}_i \cdot \nabla \right) f_i^{\mathrm{eq}}
+ \mathcal{O}(\Delta t^2).
\label{fneq_cont}
\end{equation}
The Chapman--Enskog expansion establishes the link between the microscopic LBE and the
macroscopic diffusion equation: the $\mathcal{O}(1)$ balance gives conservation of mass,
while the $\mathcal{O}(\Delta t)$ balance yields the diffusion equation with diffusivity
$d = c_s^2(\tau - 1/2)$. To avoid explicit derivative evaluations, the material
derivative along characteristics in Eq.~(\ref{fneq_cont}) is approximated by a
first-order backward difference:
\begin{equation}
\left( \partial_t + \mathbf{c}_i \cdot \nabla \right) f_i^{\mathrm{eq}}
\approx
\frac{
f_i^{\mathrm{eq}}(\mathbf{x}, t)
- f_i^{\mathrm{eq}}(\mathbf{x}-\mathbf{c}_i \Delta t,\, t-\Delta t)
}{\Delta t}.
\label{char_fd}
\end{equation}
Substituting Eq.~(\ref{char_fd}) into Eq.~(\ref{fneq_cont}) and using $f_i^{\mathrm{eq}}
= w_i\rho$,
\begin{equation}
f_i(\mathbf{x}, t)
= (1-\tau)\, w_i\rho(\mathbf{x}, t)
+ \tau\, w_i\rho(\mathbf{x}-\mathbf{c}_i \Delta t,\, t-\Delta t)
+ \mathcal{O}(\Delta t^2).
\label{fi_recon}
\end{equation}
Summing over all directions and shifting $t-\Delta t \to t$ gives the explicit
single-step update
\begin{widetext}
\begin{equation}
\rho(\mathbf{x}, t+\Delta t)
= \underbrace{\sum_i w_i\,\rho(\mathbf{x}-\mathbf{c}_i \Delta t,\, t)}_{\rho_b}
+ (\tau - 1)
\left[
\underbrace{\sum_i w_i\,\rho(\mathbf{x}+\mathbf{c}_i \Delta t,\, t)}_{\rho_f}
- 2\underbrace{\rho(\mathbf{x}, t)}_{\rho_c}
+ \underbrace{\sum_i w_i\,\rho(\mathbf{x}-\mathbf{c}_i \Delta t,\, t)}_{\rho_b}
\right]
+ \mathcal{O}(\Delta t^2).
\label{sslbm_clean}
\end{equation}
\end{widetext}
Including the reaction source term explicitly, the final SSLBM update for compartment $C$
is
\begin{align}
\rho^C(\mathbf{x}, t+\Delta t)
= &\rho_b^C
+ (\tau^C - 1)\!\left( \rho_f^C - 2\rho_c^C + \rho_b^C \right)+ \nonumber \\
&\Psi^C(\mathbf{x}, t)\Delta t,
\label{sslbm_final_clean}
\end{align}
where
\begin{align}
\rho_f^C &= \sum_i w_i\,\rho^C(\mathbf{x}+\mathbf{c}_i \Delta t,\, t), \label{rhof}\\
\rho_b^C &= \sum_i w_i\,\rho^C(\mathbf{x}-\mathbf{c}_i \Delta t,\, t), \label{rhob}\\
\rho_c^C &= \rho^C(\mathbf{x}, t). \label{rhoc}
\end{align}
\indent The combination $\rho_f^C - 2\rho_c^C + \rho_b^C$ is algebraically equivalent to a second-order centred
finite difference approximation of the Laplacian $\Delta t^2 c_s^2 \nabla^2 \rho^C$ (as
shown explicitly in the consistency analysis below). This structural similarity to a
standard centred FDM is not coincidental: both schemes use the same nearest-neighbour
stencil. However, the SSLBM differs from a standard FDM in three important respects.
First, the diffusivity $d^C$ is encoded implicitly through the relaxation parameter
$\tau^C$ via Eq.~(\ref{diffusivity}), rather than appearing as an explicit coefficient in
the stencil. Second, on a D2Q5 lattice the weighted averages $\rho_f^C$ and $\rho_b^C$
include contributions from all five stencil directions (the rest direction and the four
axis-aligned neighbours), providing an isotropic discretisation of the Laplacian. Third,
and most importantly, the SSLBM inherits its derivation and physical interpretation from
the underlying kinetic description: the update rule arises from the Chapman--Enskog
expansion of the LBE, preserving the connection between the mesoscopic PDF dynamics and
the macroscopic diffusion equation.\\
\indent The SSLBM~(\ref{sslbm_final_clean}) applies to each of the four mobile SEIRD
compartments, supplemented by the ODE
\begin{equation}
    \partial_t \rho^D = \phi \gamma \rho^I
    \label{rhoD}
\end{equation}
for the deceased. Each grid point is processed exactly once per time step, and the method stores only $10M$ values, matching the SLBM memory footprint and requiring less than half the storage of the BGK LBM.\\

\subsubsection{Relation to finite-difference methods}

At first sight, the SSLBM update~(\ref{sslbm_final_clean}) bears a strong resemblance to a standard second-order centred finite-difference (FDM) discretisation of the diffusion operator. In particular, the combination
\[
\rho_f - 2\rho_c + \rho_b
\]
corresponds to a nearest-neighbour stencil that approximates the Laplacian. This raises the natural question of whether the SSLBM is, in essence, equivalent to a classical FDM scheme.

While there is indeed a close algebraic connection at the level of the discrete operator, the two approaches differ in several important and fundamental ways.

\paragraph{Kinetic versus phenomenological derivation.}
The SSLBM is derived from the lattice Boltzmann equation through a Chapman--Enskog expansion, and therefore inherits a kinetic interpretation in which diffusion arises from an underlying particle transport process. In contrast, finite-difference methods are constructed directly at the macroscopic level as discretisations of differential operators. As a result, the SSLBM preserves a link between mesoscopic dynamics and macroscopic behaviour, which is absent in standard FDM formulations.

\paragraph{Implicit encoding of transport coefficients.}
In the SSLBM, the diffusion coefficient is not introduced explicitly in the stencil, but is instead controlled through the relaxation parameter $\tau$ via Eq.~(\ref{diffusivity}). This parameterisation is inherited from the BGK collision operator and allows transport properties to be modified without altering the discrete stencil. In contrast, FDM schemes incorporate diffusivity directly as a multiplicative coefficient in the discretised Laplacian.

\paragraph{Structural equivalence and spectral properties.}
Although the SSLBM diffusion operator is algebraically equivalent to a second-order centred finite-difference Laplacian on the same stencil, its derivation imposes specific weights and symmetries inherited from the underlying lattice. As shown in the Fourier analysis, the resulting operator has a well-defined lattice symbol and associated spectral properties. This provides a natural framework for analysing anisotropy and stability in terms of the lattice structure, rather than purely in terms of truncation error.

\paragraph{Extension to complex transport models.}
A key advantage of the SSLBM formulation is its direct connection to the broader lattice Boltzmann framework. This makes it straightforward to extend the method to more complex transport processes, including nonlinear diffusion, anisotropic transport, and coupled multi-physics problems, by modifying the underlying kinetic description. In contrast, extending FDM schemes to such settings often requires ad hoc modifications of the discretisation.

\paragraph{Computational structure.}
From an implementation perspective, the SSLBM retains the locality and stencil-based structure of lattice Boltzmann methods, with a single-pass update per time step. While FDM schemes share similar locality properties, the SSLBM benefits from a formulation that naturally aligns with data-parallel architectures and inherits optimisation strategies developed for LBM solvers.

\noindent
In summary, although the SSLBM update can be written in a form that resembles a centred finite-difference scheme, it should be viewed as a macroscopic reformulation of a kinetic method rather than a purely phenomenological discretisation. This distinction is particularly relevant when considering extensions beyond simple diffusion, where the kinetic origin of the method provides additional modelling flexibility. In this sense, the SSLBM may be interpreted as a bridge between kinetic schemes and classical finite-difference methods, combining the interpretability of the former with the efficiency of the latter.

\subsubsection{Scope of analysis}
We now formally assess the consistency, stability, anisotropy error, and mass-conservation properties of the SSLBM. The analysis is carried out for a generic scalar field $\rho$ satisfying Eq.~(\ref{generic_pde}), with the SEIRD-specific results following as a
special case.
\paragraph{Consistency.}
We consider a single compartment with purely diffusive dynamics ($\Psi = 0$) on a uniform
D2Q5 lattice with $\Delta x = \Delta y = \Delta t = 1$ (in lattice units). Expanding the forward and backward
stencil contributions via Taylor series,
\begin{align} 
\rho(\mathbf{x} \pm \mathbf{c}_i \Delta t, t) &= \rho \pm \Delta t \, \mathbf{c}_i \cdot \nabla \rho + \frac{\Delta t^2}{2} (\mathbf{c}_i \cdot \nabla)^2 \rho \nonumber \\ 
&\pm \frac{\Delta t^3}{6} (\mathbf{c}_i \cdot \nabla)^3 \rho + \mathcal{O}(\Delta t^4). \end{align} 
Using the lattice symmetry relations 
\begin{align} \sum_i w_i &= 1, & \sum_i w_i \mathbf{c}_i &= 0, & \sum_i w_i c_{i\alpha} c_{i\beta} &= c_s^2 \delta_{\alpha\beta}, 
\end{align} 
we obtain 
\begin{align} \rho_f &= \rho + \frac{\Delta t^2}{2} c_s^2 \nabla^2 \rho + \mathcal{O}(\Delta t^4), \\ \rho_b &= \rho + \frac{\Delta t^2}{2} c_s^2 \nabla^2 \rho + \mathcal{O}(\Delta t^4), 
\end{align} 
and thus 
\begin{equation} \rho_f - 2\rho_c + \rho_b = \Delta t^2 c_s^2 \nabla^2 \rho + \mathcal{O}(\Delta t^4). 
\end{equation} 
Expanding the left-hand side in time yields 
\begin{equation} \rho(\mathbf{x}, t+\Delta t) = \rho + \Delta t \, \partial_t \rho + \frac{\Delta t^2}{2} \partial_t^2 \rho + \mathcal{O}(\Delta t^3), 
\end{equation} 
so that the SSLBM recovers the macroscopic diffusion equation 
\begin{equation} \partial_t \rho = d \nabla^2 \rho + \mathcal{O}(\Delta t^2), \quad d = c_s^2 (\tau - 1/2), 
\end{equation} 
demonstrating second-order spatial accuracy.

\paragraph{Stability: purely diffusive case ($\Psi = 0$).}
We apply a von~Neumann analysis by substituting the Fourier mode
\begin{equation}
\rho(\mathbf{x}, t) = \hat{\rho}\, e^{i(k_x x + k_y y)} G^n,
\end{equation}
where $\hat{\rho}$ denotes the (complex) Fourier coefficient of the mode with wavevector $\mathbf{k}$ and $G$ is the amplification factor. Define the lattice symbol
\begin{equation}
\Lambda(\mathbf{k}) = \sum_{i=0}^{4} w_i e^{i \mathbf{k}\cdot \mathbf{c}_i}.
\end{equation}
For the D2Q5 lattice with weights $w_0 = 1/3$ and $w_{1\ldots4} = 1/6$, this evaluates to
\begin{equation}
\Lambda(\mathbf{k}) = \frac{1}{3} + \frac{1}{3}\big(\cos k_x + \cos k_y\big),
\label{Lambda_D2Q5}
\end{equation}
which is purely real due to lattice symmetry. Consequently,
\begin{equation}
\rho_f = \Lambda\, \rho, \qquad
\rho_b = \Lambda\, \rho, \qquad
\rho_c = \rho.
\end{equation}
Substituting into the SSLBM update~(\ref{sslbm_clean}) yields the exact amplification factor
\begin{align}
G(\mathbf{k})
&= \Lambda(\mathbf{k}) + (\tau-1)(\Lambda(\mathbf{k}) - 2 + \Lambda(\mathbf{k})) \nonumber \\
&= \Lambda(\mathbf{k}) + 2(\tau-1)(\Lambda(\mathbf{k}) - 1) \nonumber \\
&= 1 - 2(\tau-1)\big(1 - \Lambda(\mathbf{k})\big).
\label{G_exact}
\end{align}
A key identity linking the lattice symbol to the discrete Laplacian is
\begin{equation}
1 - \Lambda(\mathbf{k})
= \frac{1}{3}\big[(1-\cos k_x) + (1-\cos k_y)\big],
\label{Lambda_identity}
\end{equation}
which shows explicitly that $1 - \Lambda(\mathbf{k}) \ge 0$ for all wave numbers. This identity shows that the SSLBM diffusion operator is spectrally equivalent to the standard second-order finite-difference Laplacian on the same stencil, and makes explicit that $1 - \Lambda(\mathbf{k}) \ge 0$ for all wave numbers.\\
\indent The amplification factor is therefore real and satisfies
\begin{equation}
0 \le G(\mathbf{k}) \le 1 \quad \text{for all } \mathbf{k}
\end{equation}
provided $\tau \ge 1/2$. This recovers the standard BGK LBM stability condition.

\paragraph{Stability: reaction-diffusion case ($\Psi \ne 0$).}
The SEIRD source terms~(\ref{sources}) are nonlinear, so a direct von~Neumann analysis of
the fully coupled system is not feasible in general. We instead linearise about the
disease-free equilibrium (DFE), the natural reference state for epidemic onset:
\begin{equation}
\rho^S_0 = \rho^{N_c}, \qquad \rho^E_0 = \rho^I_0 = \rho^R_0 = \rho^D_0 = 0.
\label{DFE}
\end{equation}
Writing $\rho^C = \rho^C_0 + \delta\rho^C$ and using $\rho^S_0/\rho^{N_c} = 1$, the
linearised source terms are
\begin{eqnarray}
\delta\Psi^{S} &=& -\beta\,\delta\rho^I, \label{linS}\\
\delta\Psi^{E} &=& +\beta\,\delta\rho^I - \alpha\,\delta\rho^E, \label{linE}\\
\delta\Psi^{I} &=& +\alpha\,\delta\rho^E - \gamma\,\delta\rho^I, \label{linI}\\
\delta\Psi^{R} &=& +(1-\phi)\gamma\,\delta\rho^I, \label{linR}
\end{eqnarray}
where $\delta \Psi^C$ and $\delta \rho^C$ are small perturbations around the disease-free equilibrium. Note that $\delta\rho^S$ does not appear in $\delta\Psi^E$, $\delta\Psi^I$, or
$\delta\Psi^R$: the susceptible compartment decouples from the infection dynamics at
leading order near the DFE. The linearised SSLBM update for each perturbation reads
\begin{equation}
\widehat{\delta\rho}^C(t + \Delta t)
= G_{\mathrm{diff}}(\mathbf{k})\,\widehat{\delta\rho}^C(t)
+ \Delta t \sum_{C'} J_{CC'}\,\widehat{\delta\rho}^{C'}(t),
\label{linearised_update}
\end{equation}
where $\widehat{\delta\rho}$ denotes the Fourier amplitude of the perturbation, $G_{\mathrm{diff}}(\mathbf{k}) = 1 - 2(\tau-1)\big(1 - \Lambda(\mathbf{k})\big)$
is the amplification factor of the purely diffusive SSLBM scheme and $J$ is the $4\times 4$ Jacobian of the linearised source terms with respect to $(\delta\rho^S, \delta\rho^E, \delta\rho^I, \delta\rho^R)$:
\begin{equation}
J =
\begin{pmatrix}
0 & 0 & -\beta & 0 \\
0 & -\alpha & +\beta & 0 \\
0 & +\alpha & -\gamma & 0 \\
0 & 0 & (1-\phi)\gamma & 0
\end{pmatrix}.
\label{jacobian}
\end{equation}
The full amplification operator is
\begin{equation}
\mathcal{G}(\mathbf{k}) = G_{\mathrm{diff}}(\mathbf{k})\,\mathbf{I} + \Delta t\,J,
\label{amplification_matrix}
\end{equation}
with eigenvalues $\mu_j = G_{\mathrm{diff}} + \Delta t\,\lambda_j$, $j = 1,\ldots,4$,
where $\lambda_j$ are the eigenvalues of $J$. Two of these are zero (from the decoupled
$S$ and $R$ rows), so their modes satisfy $|\mu| = |G_{\mathrm{diff}}| \le 1$ whenever
$\tau \ge 1/2$. The dynamically significant eigenvalues come from the $E$--$I$ sub-block
\begin{equation}
J_{EI} =
\begin{pmatrix}
-\alpha & \beta \\
+\alpha & -\gamma
\end{pmatrix},
\label{JEI}
\end{equation}
whose characteristic polynomial is
\begin{equation}
\lambda^2 + (\alpha + \gamma)\lambda + \alpha(\gamma - \beta) = 0,
\end{equation}
giving the two real eigenvalues
\begin{equation}
\lambda_{1,2} = \frac{-(\alpha+\gamma) \pm \sqrt{(\alpha - \gamma)^2 + 4\alpha\beta}}{2}.
\label{eigenvalues}
\end{equation}
The larger eigenvalue $\lambda_1$ (with the $+$ sign) governs the stability of the
reaction subsystem. It satisfies $\lambda_1 < 0$ if and only if $\gamma > \beta$,
which is equivalent to
\begin{equation}
R_0 \equiv \frac{\beta}{\gamma} < 1,
\label{R0_condition}
\end{equation}
recovering the classical epidemic threshold. We remark that for a spatially
homogeneous system the basic reproduction number takes the simple form $R_0 =
\beta/\gamma$. For the spatially extended PDE model, the definition of $R_0$ is more
subtle and typically requires a next-generation matrix
analysis~\cite{van2002reproduction,diekmann2010construction}; however, near a spatially
uniform DFE, the spatially homogeneous $R_0$ remains a useful indicator of the stability
of the DFE.

When $R_0 < 1$, both $\lambda_{1,2} < 0$, so $\mathcal{R}(J) = |\lambda_1|$, and the combined
stability condition $|\mu_j| \le 1$ requires
\begin{equation}
\Delta t \le \frac{2}{\mathcal{R}(J)},
\label{dt_condition}
\end{equation}
$\mathcal{R}(J)$ being the spectral radius of the matrix $J$. Since $\lambda_1 < 0$ when $R_0 < 1$, we have $\mathcal{R}(J) = -\lambda_1$. Substituting
from Eq.~(\ref{eigenvalues}),
\begin{equation}
\mathcal{R}(J) = \frac{(\alpha+\gamma) - \sqrt{(\alpha - \gamma)^2 + 4\alpha\beta}}{2},
\end{equation}
which is strictly positive when $R_0 < 1$ since $(\alpha+\gamma)^2 > (\alpha -
\gamma)^2 + 4\alpha\beta$ in that regime. The time-step condition $\Delta t \le
2/\mathcal{R}(J)$ therefore becomes
\begin{equation}
\Delta t \le \frac{4}{(\alpha + \gamma) - \sqrt{(\alpha-\gamma)^2 + 4\alpha\beta}}.
\label{dt_explicit}
\end{equation}
This condition supplements the diffusion stability requirement $\tau \ge 1/2$ and,
together with it, constitutes the complete stability requirements for the SSLBM applied
to the SEIRD model in the linearised regime near the DFE.

\indent We remark that the analysis above is performed in the linearised regime and does
not provide rigorous guarantees for the full nonlinear system. The reaction terms are
treated explicitly, which is standard for non-stiff reaction-diffusion systems. For the
epidemiological parameters considered here this condition is satisfied, and the numerical
results confirm the stability of the scheme throughout the full nonlinear simulation.

\paragraph{Anisotropy of the SSLBM diffusion operator}

The SSLBM diffusion operator can be expressed in Fourier space via its lattice symbol,
\begin{equation}
\Lambda_{\text{SSLBM}}(\mathbf{k}) = \sum_i w_i \, e^{i \mathbf{k}\cdot \mathbf{c}_i},
\end{equation}
where $\mathbf{c}_i$ and $w_i$ are the lattice directions and weights. Expanding for small wavenumbers $(k_x, k_y)$ on the D2Q5 cross stencil gives
\begin{align}
1 - \Lambda_{\text{D2Q5}}(\mathbf{k}) &= \frac{1}{6} (k_x^2 + k_y^2)
- \frac{1}{72} (k_x^4 + k_y^4) + \mathcal{O}(|\mathbf{k}|^6) \nonumber \\
&= \frac{1}{6} |\mathbf{k}|^2 - \frac{1}{72} |\mathbf{k}|^4 + \frac{1}{36} k_x^2 k_y^2 + \mathcal{O}(|\mathbf{k}|^6).
\end{align}
The leading-order term $\frac{1}{6}|\mathbf{k}|^2$ is isotropic, recovering the standard Laplacian. The fourth-order term contains the mixed $k_x^2 k_y^2$ contribution, which is the \emph{anisotropy error} of the cross stencil. This term vanishes along coordinate axes and diagonals but introduces a directional bias for intermediate angles.

For comparison, the D2Q9 lattice (9-point stencil) has symbol
\begin{align}
\Lambda_{\text{D2Q9}}(\mathbf{k}) = &\frac{4}{9} + \frac{1}{9}(\cos k_x + \cos k_y) + \nonumber \\
&\frac{1}{36} \left[ \cos(k_x+k_y) + \cos(k_x-k_y) \right].
\end{align}
Its small-wavenumber expansion reads
\begin{equation}
1 - \Lambda_{\text{D2Q9}}(\mathbf{k}) = \frac{1}{6} (k_x^2 + k_y^2) - \frac{1}{72} (k_x^2 + k_y^2)^2 + \mathcal{O}(|\mathbf{k}|^6),
\end{equation}
which is isotropic up to fourth order, with no $k_x^2 k_y^2$ term. Hence, D2Q9 reduces anisotropy at high wavenumbers, providing a more accurate spectral approximation of the Laplacian. This comparison highlights that:
\begin{enumerate}
    \item D2Q5: second-order isotropic, fourth-order anisotropic; simplest stencil, minimal memory.
    \item D2Q9: second-order isotropic, fourth-order isotropic; better isotropy at the cost of extra storage and computations.
\end{enumerate}
This also shows that the SSLBM diffusion operator is spectrally equivalent to the standard second-order finite-difference Laplacian on the same stencil. The anisotropy error therefore arises purely from the stencil geometry, not from the kinetic derivation.

\paragraph{Mass conservation.}
We now show that the SSLBM exactly conserves the total population $N = \sum_C
\sum_{\mathbf{x}} \rho^C(\mathbf{x},t)$ at the discrete level. Summing
Eq.~(\ref{sslbm_final_clean}) over all grid points $\mathbf{x}$ for a given compartment
$C$, the diffusion terms $\rho_f^C - 2\rho_c^C + \rho_b^C$ vanish by periodicity (or by
homogeneous Neumann boundary conditions), since they represent a discretised Laplacian
whose spatial sum is zero. Therefore
\begin{equation}
\sum_{\mathbf{x}} \rho^C(\mathbf{x}, t+\Delta t) = \sum_{\mathbf{x}} \rho^C(\mathbf{x}, t)
+ \Delta t \sum_{\mathbf{x}} \Psi^C(\mathbf{x}, t).
\end{equation}
Summing now over all mobile compartments $C \in \{S, E, I, R\}$ and using the ODE for
$\rho^D$, the net reactive contributions telescope:
\begin{align}
\sum_C \sum_{\mathbf{x}} \Psi^C
&= \sum_{\mathbf{x}} \Bigg[
- \frac{\beta \rho^S \rho^I}{\rho^{N_c}}
+ \frac{\beta \rho^S \rho^I}{\rho^{N_c}}
- \alpha \rho^E + \alpha \rho^E \nonumber \\
&\qquad\qquad
- \gamma \rho^I
+ (1-\phi)\gamma \rho^I
+ \phi \gamma \rho^I
\Bigg] = 0.
\end{align}
where the last two terms come from the $R$ and $D$ equations respectively. All reactive
terms cancel exactly, yielding
\begin{equation}
\sum_C \sum_{\mathbf{x}} \rho^C(\mathbf{x}, t+\Delta t)
= \sum_C \sum_{\mathbf{x}} \rho^C(\mathbf{x}, t).
\label{mass_conservation}
\end{equation}
The SSLBM therefore \emph{exactly conserves} the total population at the discrete level,
an important property for epidemic models in which the total population must remain
constant in the absence of births and non-disease deaths.

% -----------------------------------------------------------------------
% RESULTS AND VALIDATION
% -----------------------------------------------------------------------
\section{Results}
We validate the SSLBM against a high-order finite difference (FDM) reference solution of Eqs.~(\ref{EqSEIRD_S}--\ref{EqSEIRD_D}). The reference solver employs a fully explicit fourth-order Runge--Kutta time integration combined with fourth-order centred spatial discretisation. Nonlinear reaction terms are treated explicitly at each stage.\\
\indent The test case considers a two-dimensional domain of size $200 \times 200\,\mathrm{km}^2$ with total population $N = 5 \times 10^6$. Epidemiological parameters are chosen as $\gamma = (5\,\mathrm{days})^{-1}$, $\beta = R_0 \gamma$, $\alpha = (7\,\mathrm{days})^{-1}$, and $\phi = 0.3$. Diffusion coefficients are $d^{S} = d^{E} = d^{R} = 4.35 \times 10^{-2}\,\mathrm{km}^2\,\mathrm{day}^{-1}$ and $d^{I} = 10^{-4}\,\mathrm{km}^2\,\mathrm{day}^{-1}$, reflecting reduced mobility of infectious individuals~\cite{VIGUERIE2021106617}.\\
\indent The initial condition consists of a nearly homogeneous population with a prescribed fraction
$\chi$ of exposed individuals, perturbed by a small random field $\varepsilon(\mathbf{x})$:
\begin{eqnarray}
\rho^N(\mathbf{x},0) &=& N/A, \nonumber \\
\rho^E(\mathbf{x},0) &=& \chi N/A + \varepsilon(\mathbf{x}), \nonumber \\
\rho^S(\mathbf{x},0) &=& \rho^N(\mathbf{x},0) - \rho^E(\mathbf{x},0),
\end{eqnarray}
with $\rho^I = \rho^R = \rho^D = 0$. Simulations are performed with $\Delta x = 1\,\mathrm{km}$ and
$\Delta t = 10^{-2}\,\mathrm{days}$.

\subsection{Accuracy}
\begin{figure*}[!htbp]
\subfigure[0.01]{\includegraphics[width=0.49\textwidth]{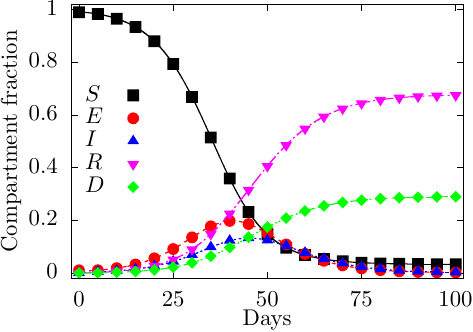}}
\subfigure[0.05]{\includegraphics[width=0.49\textwidth]{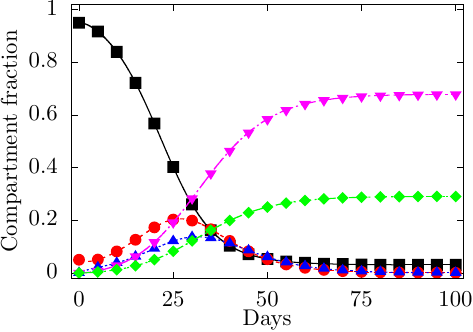}}
\subfigure[0.10]{\includegraphics[width=0.49\textwidth]{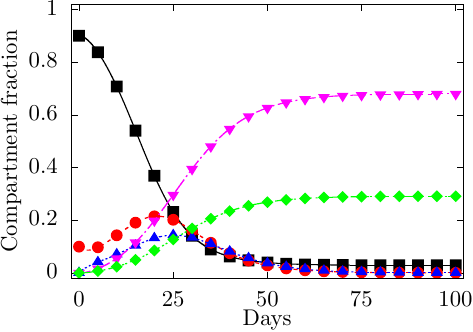}}
\subfigure[0.20]{\includegraphics[width=0.49\textwidth]{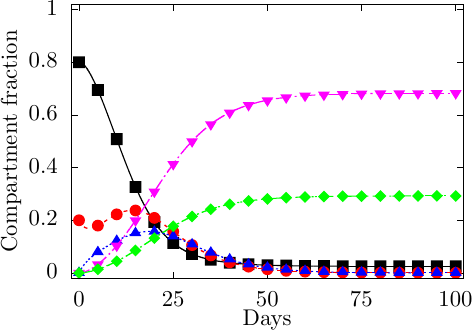}}
\caption{Temporal evolution of SEIRD compartments computed with SSLBM (lines) and FDM (symbols)
for increasing initial exposure fraction $\chi$. Excellent agreement is observed across all compartments.}
\label{Figure1}
\end{figure*}

\begin{figure}[!htbp]
\subfigure[$R_0=6$]{\includegraphics[width=0.49\textwidth]{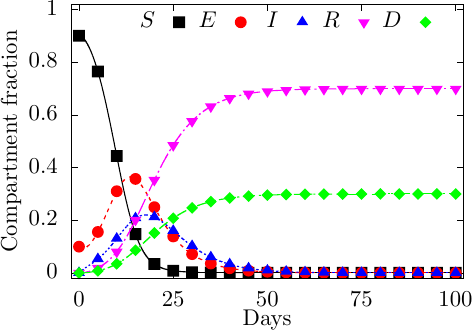}}
\subfigure[$R_0=12$]{\includegraphics[width=0.49\textwidth]{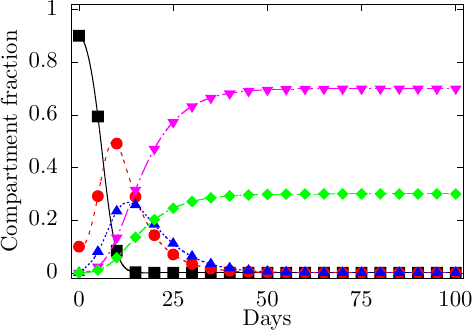}}
\subfigure[$R_0=18$]{\includegraphics[width=0.49\textwidth]{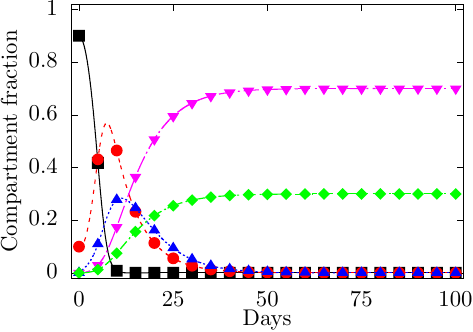}}
\caption{Temporal evolution of SEIRD compartments for increasing reproduction number $R_0$.
SSLBM (symbols) and FDM (lines) remain in excellent agreement even in highly super-critical regimes.}
\label{Figure2}
\end{figure}

Figure~\ref{Figure1} compares the temporal evolution of all compartments predicted by the SSLBM and the FDM for $\chi \in \{0.01, 0.05, 0.10, 0.20\}$ at $R_0=3$. Across all cases, the SSLBM reproduces the full epidemic dynamics with excellent accuracy, including the timing and amplitude of the infection peak and the subsequent decay of the epidemic wave. No significant phase shift or systematic bias is observed, even for larger initial perturbations where nonlinear effects are stronger.
\\
\indent Table~\ref{tab:chi_combined} quantifies the influence of the initial exposure level $\chi$ on the numerical accuracy. Across all compartments and norms, the SSLBM consistently outperforms the BGK formulation. A clear trend emerges: the relative improvement tends to increase with with $\chi$.
For weak perturbations ($\chi=0.01$), the SSLBM reduces the error by approximately a factor of $1.5$--$1.7$. As $\chi$ increases, the improvement becomes progressively stronger, reaching factors of $2$--$3$ for intermediate regimes and exceeding $4$ in the $L_\infty$ norm for $\chi=0.20$. This behaviour indicates that the SSLBM becomes increasingly advantageous in regimes characterised by stronger diffusion-driven spatial gradients and enhanced nonlinear coupling. The particularly large gain in the $L_\infty$ norm further suggests improved control of localised extrema, where BGK formulations tend to produce larger pointwise deviations. Importantly, the improvement is consistent across all compartments, with slightly stronger gains observed in the exposed and infected populations, where steep gradients are most pronounced. Despite these differences, all errors remain below $0.1\%$, confirming that both schemes operate
in a highly accurate regime at the chosen resolution.
\begin{table*}[!htbp]
\centering
\setlength{\tabcolsep}{5pt}
\renewcommand{\arraystretch}{1.15}
\begin{tabular}{c|c|ccc|ccc}
%\toprule
\hline
\hline
\multirow{ 2}{*}{$\chi$} & \multirow{ 2}{*}{Comp.}
& \multicolumn{3}{c|}{$L_2$-norm (\%)} 
& \multicolumn{3}{c}{$L_\infty$-norm (\%)} \\
\cline{3-5} \cline{6-8}
 &  & SSLBM & BGK & Ratio 
    & SSLBM & BGK & Ratio \\
%\midrule
\hline
\multirow{4}{*}{0.01}
 & $S$   & 0.0364 & 0.0624 & 1.71 & 0.0482 & 0.0799 & 1.66 \\
 & $E$   & 0.0719 & 0.1208 & 1.68 & 0.0755 & 0.1154 & 1.53 \\
 & $I$   & 0.0711 & 0.1158 & 1.63 & 0.0750 & 0.1097 & 1.46 \\
 & $R,D$ & 0.0284 & 0.0475 & 1.67 & 0.0431 & 0.0696 & 1.61 \\
%\midrule
\hline
\multirow{4}{*}{0.05}
 & $S$   & 0.0241 & 0.0603 & 2.50 & 0.0252 & 0.0603 & 2.39 \\
 & $E$   & 0.0384 & 0.0856 & 2.23 & 0.0468 & 0.0856 & 1.83 \\
 & $I$   & 0.0413 & 0.0836 & 2.02 & 0.0511 & 0.0836 & 1.64 \\
 & $R,D$ & 0.0137 & 0.0508 & 3.71 & 0.0233 & 0.0508 & 2.18 \\
%\midrule
\hline
\multirow{4}{*}{0.10}
 & $S$   & 0.0204 & 0.0591 & 2.90 & 0.0162 & 0.0549 & 3.39 \\
 & $E$   & 0.0303 & 0.0692 & 2.28 & 0.0369 & 0.0740 & 2.01 \\
 & $I$   & 0.0342 & 0.0717 & 2.10 & 0.0436 & 0.0994 & 2.28 \\
 & $R,D$ & 0.0104 & 0.0239 & 2.30 & 0.0158 & 0.0441 & 2.79 \\
%\midrule
\hline
\multirow{4}{*}{0.20}
 & $S$   & 0.0234 & 0.0696 & 2.97 & 0.0121 & 0.0609 & 5.03 \\
 & $E$   & 0.0291 & 0.0601 & 2.07 & 0.0345 & 0.1168 & 3.39 \\
 & $I$   & 0.0332 & 0.0725 & 2.18 & 0.0384 & 0.1811 & 4.72 \\
 & $R,D$ & 0.0094 & 0.0212 & 2.26 & 0.0135 & 0.0424 & 3.14 \\
%\bottomrule
\hline
\hline
\end{tabular}
\caption{Percentage relative $L_2$- and $L_\infty$-norm errors for SSLBM and BGK LBM across increasing $\chi$. The ratio (BGK/SSLBM) quantifies the accuracy gain of SSLBM. While the improvement is modest for small $\chi$, it increases significantly with $\chi$, reaching factors above $4$ in the $L_\infty$ norm. The $R$ and $D$ compartments exhibit identical errors and are reported jointly.}
\label{tab:chi_combined}
\end{table*}
\\
\indent We further investigate more demanding epidemiological regimes by increasing the basic reproduction number to $R_0 = 6, 12, 18$ at fixed $\chi = 0.1$. Figure~\ref{Figure2} shows that increasing $R_0$ leads to faster epidemic growth, sharper infection peaks, and stronger nonlinear coupling between compartments. Despite these increasingly stiff dynamics, the SSLBM remains in excellent agreement with the reference solution across all compartments,
without any sign of instability or oscillatory artefacts, even in the strongly super-critical regime $R_0=18$.
\\
\indent Table~\ref{tab:combined_ratio} confirms these observations quantitatively. The SSLBM consistently reduces the error by a factor of approximately $2$--$2.5$ across all compartments and norms. Notably, this gain is largely insensitive to the epidemiological regime, indicating that the method maintains its accuracy advantage even as nonlinear effects become more pronounced. While the absolute error increases with $R_0$, reflecting sharper spatial and temporal gradients, the relative improvement of the SSLBM remains stable. This highlights the robustness of the proposed formulation under increasingly stiff reaction–diffusion dynamics.
\begin{table*}[!htbp]
\centering
\setlength{\tabcolsep}{5pt}
\renewcommand{\arraystretch}{1.15}
\begin{tabular}{c|c|ccc|ccc}
%\toprule
\hline
\hline
\multirow{ 2}{*}{$R_0$} & \multirow{ 2}{*}{Comp.}
& \multicolumn{3}{c|}{$L_2$-norm (\%)} 
& \multicolumn{3}{c}{$L_\infty$-norm (\%)} \\
\cline{3-5} \cline{6-8}
 &  & SSLBM & BGK & Ratio 
    & SSLBM & BGK & Ratio \\
%\midrule
\hline
\multirow{4}{*}{6}
 & $S$   & 0.0504 & 0.1345 & 2.67 & 0.0527 & 0.1308 & 2.48 \\
 & $E$   & 0.0584 & 0.1361 & 2.33 & 0.0670 & 0.1405 & 2.10 \\
 & $I$   & 0.0602 & 0.1252 & 2.08 & 0.0711 & 0.1325 & 1.86 \\
 & $R,D$ & 0.0139 & 0.0319 & 2.29 & 0.0366 & 0.0785 & 2.14 \\
%\midrule
\hline
\multirow{4}{*}{12}
 & $S$   & 0.1019 & 0.2472 & 2.43 & 0.1091 & 0.2484 & 2.28 \\
 & $E$   & 0.0937 & 0.2180 & 2.33 & 0.1083 & 0.2577 & 2.38 \\
 & $I$   & 0.0853 & 0.1778 & 2.08 & 0.1019 & 0.2291 & 2.25 \\
 & $R,D$ & 0.0174 & 0.0377 & 2.17 & 0.0536 & 0.1060 & 1.98 \\
%\midrule
\hline
\multirow{4}{*}{18}
 & $S$   & 0.1429 & 0.3333 & 2.33 & 0.1544 & 0.3398 & 2.20 \\
 & $E$   & 0.1181 & 0.2696 & 2.28 & 0.1524 & 0.3466 & 2.27 \\
 & $I$   & 0.0993 & 0.2040 & 2.05 & 0.1370 & 0.2931 & 2.14 \\
 & $R,D$ & 0.0190 & 0.0398 & 2.09 & 0.0608 & 0.1164 & 1.92 \\
%\bottomrule
\hline
\hline
\end{tabular}
\caption{Percentage relative $L_2$- and $L_\infty$-norm errors for SSLBM and BGK LBM across increasing $R_0$. The ratio (BGK/SSLBM) quantifies the accuracy gain of SSLBM, which consistently reduces the error by a factor of approximately $2$--$2.5$ across all compartments and norms. The $R$ and $D$ compartments exhibit identical errors and are reported jointly.}
\label{tab:combined_ratio}
\end{table*}
\\
\indent Taken together, our present results demonstrate that the SSLBM provides a quantitatively accurate and robust approximation of the SEIRD system, while offering a clear accuracy advantage over the standard BGK formulation. Moreover, we underline that the SSLBM retains both accuracy and stability well beyond the mildly super-critical regime, making it suitable for simulating fast-spreading epidemics.

\subsection{Convergence study: diffusion limit}
\begin{figure}[htbp]
\centering
\includegraphics[width=0.49\textwidth]{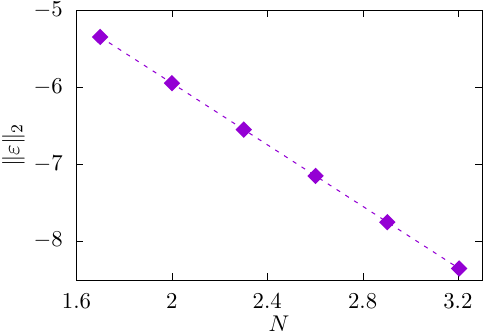}
\caption{Log--log plot of the $L_2$ error as a function of grid resolution for the diffusion test case. The dashed line indicates second-order convergence. The SSLBM results closely follow the reference slope, confirming the expected $\mathcal{O}(\Delta x^2)$ accuracy.}
\label{fig:convergence}
\end{figure}

To verify the spatial accuracy of the proposed SSLBM, we consider a pure diffusion problem obtained by disabling all reaction terms in the SEIRD system. In this limit, each compartment evolves according to a linear diffusion equation,
\begin{equation}
\partial_t \rho = D \nabla^2 \rho,
\end{equation}
which provides a clean setting for assessing numerical convergence.\\
\indent A smooth analytical solution compatible with periodic boundary conditions is prescribed,
\begin{equation}
\rho(x,y,t) = \sin\left(2\pi x\right)\sin\left(2\pi y\right)
\exp\left(-\kappa t\right),
\end{equation}
where the decay rate is given by
\begin{equation}
\kappa = 2 D (2\pi)^2.
\end{equation}
This manufactured solution ensures that both spatial and temporal derivatives are exactly known, enabling a precise error evaluation.\\
\indent The initial condition is obtained by evaluating the analytical solution at $t=0$, and periodic boundary conditions are applied in both spatial directions. Simulations are performed on a sequence of progressively refined grids. To isolate spatial errors, the time step is scaled as $\Delta t \sim \Delta x^2$, ensuring consistency with diffusive scaling.\\
\indent The numerical error is quantified using the discrete $L_2$ norm,
\begin{equation}
\|\varepsilon\|_2 =
\left[
\frac{1}{N}\sum_{i=1}^{N}
\left(\rho_i^{\text{num}} - \rho_i^{\text{exact}}\right)^2
\right]^{1/2},
\end{equation}
where $N$ is the total number of grid points. The convergence rate $p$
is estimated as
\begin{equation}
p = \log\left(\frac{\|\varepsilon_h\|_2}{\|\varepsilon_{h/2}\|_2}\right),
\end{equation}
$h$ being a certain grid resolution.\\
\indent Table~\ref{tab:convergence} reports the $L_2$ error for a sequence of uniform grid refinements together with the corresponding convergence rates. The results show a systematic reduction of the error by approximately a factor of four when the grid spacing is halved, indicating second-order accuracy.
\begin{table}[htbp]
\centering
\caption{Grid convergence study for the diffusion problem. The $L_2$ error is reported for increasing spatial resolution.}
\label{tab:convergence}
\begin{tabular}{c|c}
\hline
\hline
Grid size & $L_2$ error\\
\hline
$50 \times 50$   & $4.53 \times 10^{-6}$\\
$100 \times 100$ & $1.13 \times 10^{-6}$\\
$200 \times 200$ & $2.83 \times 10^{-7}$\\
$400 \times 400$ & $7.09 \times 10^{-8}$\\
$800 \times 400$ & $1.77 \times 10^{-8}$\\
$1600 \times 400$ & $4.43 \times 10^{-8}$ \\
\hline
\hline
\end{tabular}
\end{table}
\\
\indent This behaviour is further illustrated in Fig.~\ref{fig:convergence}, which presents the error as a function of the grid resolution in a log--log representation. The numerical results align closely with a reference slope of two, confirming the expected $\mathcal{O}(\Delta x^2)$
scaling of the method.

%% ---------------------------------------------------------------
%% Drop-in subsection: place after \subsection{Computational cost}
%% (or wherever appropriate in your Results section)
%% Requires: \usepackage{multirow,booktabs} already in preamble
%% Figure placeholders: replace the \fbox stubs with your actual
%%   \includegraphics calls once plots are generated.
%% ---------------------------------------------------------------

\subsection{Computational cost}
The computational efficiency of the proposed SSLBM is assessed by measuring the wall-clock runtime as a function of the total number of grid points $M$. Simulations are performed for a fixed epidemiological setting ($\chi = 0.10$, $R_0 = 3$) while systematically increasing the spatial resolution. The resulting runtimes, normalised by the fastest observed run (SSLBM at $M=200 \times 200$), are reported in Table~\ref{tab:timing} and compared across the SSLBM, BGK lattice Boltzmann, a standard second-order finite difference scheme, and the fourth-order finite difference reference solver.
\begin{table}[!htbp]
\centering
\begin{tabular}{c|c|c|c|c}
\hline\hline
Grid size & SSLBM & BGK & FDM 2nd & FDM 4th \\
\hline
$200 \times 200$  & 1.00 & 1.09 & 1.93 & 3.71 \\
$400 \times 400$  & 3.99 & 5.24 & 7.87 & 15.03 \\
$600 \times 600$  & 9.11 & 10.91 & 17.94 & 34.46 \\
$800 \times 800$  & 16.42 & 21.13 & 32.51 & 62.08 \\
$1000 \times 1000$ & 25.80 & 30.65 & 50.55 & 96.41 \\
\hline\hline
\end{tabular}
\caption{Wall-clock runtime normalised by the fastest observed run (SSLBM at $M=200 \times 200$).}
\label{tab:timing}
\end{table}

\begin{figure}[!htbp]
\includegraphics[width=0.49\textwidth]{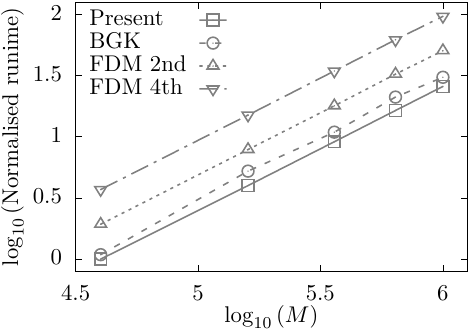}
\caption{Normalised runtime as a function of total grid points $M$: SSLBM (solid line with squares); BGK LBM (dashed line with circles); second-order FDM (dotted line with triangles); fourth-order FDM (dash-dotted line with inverted triangles). The SSLBM requires the fewest computational resources across all grid sizes tested.}
\label{Figure3}
\end{figure}

As confirmed by Figure~\ref{Figure3}, all methods exhibit a linear scaling with respect to $M$, consistent with the expected $O(M)$ complexity of explicit local update schemes. However, significant differences in prefactors are observed.\\
\indent The lattice Boltzmann formulations substantially outperform the finite difference approaches across all resolutions. In particular, the SSLBM consistently achieves the lowest runtime among all tested methods. For example, for the finest grid resolution, the SSLBM is approximately $1.2$--$1.4\times$ faster than the BGK formulation and roughly $2$--$2.5\times$ faster than the finite difference schemes.\\
\indent This performance gain is attributed to the reduced algorithmic complexity of the SSLBM, which eliminates the need for distribution functions and streaming operations. As a result, the method performs a single local update per grid node, improving memory locality and reducing memory bandwidth requirements.\\
\indent The BGK formulation, while still efficient compared to finite difference methods, incurs additional overhead due to the reconstruction of equilibrium distributions and the handling of multiple discrete velocity populations. The finite difference methods exhibit the highest computational cost, particularly for the fourth-order scheme, which requires wider stencils and additional arithmetic operations.\\
\indent Summing up, the results confirm that the SSLBM provides the most efficient computational framework among the tested approaches, combining low algorithmic complexity with favourable memory access patterns, while maintaining high numerical accuracy.

% -----------------------------------------------------------------------
% CONCLUSION
% -----------------------------------------------------------------------
\section{Conclusions}
We have introduced a single-step simplified lattice Boltzmann method for the solution of compartmental reaction--diffusion systems, with application to a spatial SEIRD epidemic model. By eliminating particle distribution functions and recasting the algorithm into a single local update, the proposed formulation achieves the key objective of combining the kinetic consistency of lattice Boltzmann methods with the compactness of macroscopic schemes.\\
\indent The analysis shows that the SSLBM recovers the correct macroscopic dynamics, inherits the key stability properties of the BGK formulation in the diffusive limit, while remaining stable under standard time-step constraints for the reactive terms, and exactly conserves the total population at the discrete level under periodic or no-flux boundary conditions. Despite its simplicity, the method retains a clear physical interpretation rooted in the underlying kinetic framework.\\
\indent Our numerical validation demonstrates that the SSLBM consistently outperforms the standard BGK lattice Boltzmann method in terms of accuracy, reducing errors by factors ranging from approximately two to five depending on the regime. The gain is particularly pronounced in strongly nonlinear scenarios and in the presence of sharp spatial gradients, where improved control of local extrema is observed. At the same time, the method remains robust across all tested epidemiological regimes, including highly super-critical regimes.\\
\indent From a computational perspective, the SSLBM achieves the lowest runtime among all methods considered, outperforming both BGK and finite difference schemes while retaining linear scaling with problem size. This efficiency stems from the single-step structure and reduced memory footprint, which improve data locality and minimise memory bandwidth requirements.\\
\indent We can conclude that the SSLBM provides a simple, second-order accurate, and computationally efficient framework for reaction--diffusion epidemic models. Its general formulation makes it immediately applicable to a broad class of compartmental systems, opening the way to efficient large-scale and high-resolution simulations of spatial epidemic dynamics. Natural directions for future work include: (i) extension to three-dimensional domains and unstructured or adaptive grids; (ii) incorporation of spatially varying and anisotropic diffusion coefficients; (iii) application to richer compartmental structures such as age-stratified or vaccine-stratified models; and (iv) exploration of the kinetic interpretation of the SSLBM formulation in the context of compartmental epidemiology, following the spirit of works on kinetic models for epidemic dynamics~\cite{zanella2023kinetic}. 

\section*{Supplemental material}
The codes used to generate the results presented in this study are freely available at \url{https://github.com/SIG-LBM-Multiphysics-Modelling/SEIRD}.

\section*{References}
\bibliographystyle{apsrev4-1}
\bibliography{bibliography}

\end{document}